\documentclass[twocolumn]{article}
\usepackage{amsmath}
\usepackage{float}
\usepackage{graphics}

\begin{document}

\title{Entropy evolution law in a laser process \thanks{{\small Work
supported by the National Natural Science Foundation of China under grant
11105133 and 11175113}} }
\author{Jun-hua Chen$^{1}$ and Hong-yi Fan$^{1}$ \thanks{
To whom correspondence should be addressed. Email: cjh@ustc.edu.cn} \\
$^{1}${\small Department of Material Science and Engineering, }\\
{\small University of Science and Technology of China,}\\
{\small Hefei, Anhui, 230026, China}}
\maketitle

\begin{abstract}
For the first time, we obtain the entropy variation law in a laser process
after finding the Kraus operator of the master equation describing the laser
process with the use of the entangled state representation. The behavior of
entropy is determined by the competition of the gain and damping in the
laser process. The photon number evolution formula is also obtained.
\end{abstract}

\section{Introduction}

Since the theoretical foundation proposed by Albert Einstein in 1917 $^{\cite%
{R1}}$ and the building of first functioning laser by Theodore H. Maiman in
1960, laser has been successfully applied in various of areas, including
laser cooling technique developed by Steven Chu et al $^{\cite{R2,R3}}$. As
one of the most important concept in physics, entropy measures the disorder
of a system. Studying the evolution of the entropy, we can get a clear
understanding of how a laser beam is created by appropriate pumping. A few
works had been done concerning the entropy exchange between a laser and its
environment $^{\cite{R4,R5}}$. However, the evolution of entropy in a laser
itself has not yet been studied before. In this work we shall derive the
entropy evolution law of a laser process. Our results explain how the
self-organization phenomenon happened in a laser.

In quantum optics theory the time evolution of laser in the lowest-order
approximation can be described by the following master equation of density
operator $^{\cite{R6,R7,R8,R9}}$
\begin{equation}
\begin{array}{c}
\frac{d\rho \left( t\right) }{dt}=g\left[ 2a^{\dagger }\rho \left( t\right)
a-aa^{\dagger }\rho \left( t\right) -\rho \left( t\right) aa^{\dagger }%
\right] \\
+\kappa \left[ 2a\rho \left( t\right) a^{\dagger }-a^{\dagger }a\rho \left(
t\right) -\rho \left( t\right) a^{\dagger }a\right] ,%
\end{array}
\label{1}
\end{equation}%
where $g$\ and $\kappa $\ are the cavity gain and the loss, respectively, $%
a^{\dagger },$ $a$ are photon creation and annihilation operator,
respectively. It is also known that the evolution due to the interaction
between a system and its environment can be ascribed to an evolution from
the initial density operator $\rho _{0}$ to $\rho \left( t\right) $%
\begin{equation}
\rho \left( t\right) =\sum_{n=0}^{\infty }M_{n}\rho _{0}M_{n}^{\dagger },
\label{2}
\end{equation}%
such an expression is named an operator-sum (Kraus) representation, $M_{n}$
is named Kraus operator. So far as our knowledge is concerned, the entropy
variation in laser-channel has not ever been reported. In this paper we
shall show how the entropy of an initial coherent state $\rho
_{0}=\left\vert z\right\rangle \left\langle z\right\vert $ (the fact that $n$%
-photon distribution in a coherent state is Poisson distribution exactly
fits the measurement result of photon distribution in a laser light) varies
in the laser process, before doing this, we should first derive the Kraus
operator by solving the master equation (\ref{1}).

Our way is introducing the two-mode entangled state
\begin{equation}
|\eta \rangle =\exp (-\frac{1}{2}|\eta |^{2}+\eta a^{\dag }-\eta ^{\ast }%
\tilde{a}^{\dag }+a^{\dag }\tilde{a}^{\dag })|0\tilde{0}\rangle ,  \label{3}
\end{equation}%
where $\tilde{a}^{\dag }$ is a fictitious mode independent of the real mode $%
a^{\dagger }$, $|\tilde{0}\rangle $ is annihilated by $\tilde{a},$ $\left[
\tilde{a},\tilde{a}^{\dag }\right] =1.$ The state $|\eta =0\rangle $
possesses the properties%
\begin{equation}
\begin{array}{c}
a|\eta =0\rangle =\tilde{a}^{\dag }|\eta =0\rangle , \\
a^{\dag }|\eta =0\rangle =\tilde{a}|\eta =0\rangle , \\
(a^{\dag }a)^{n}|\eta =0\rangle =(\tilde{a}^{\dag }\tilde{a})^{n}|\eta
=0\rangle .%
\end{array}
\label{4}
\end{equation}%
Operating the both sides of (\ref{1}) on the state $|\eta =0\rangle \equiv
\left\vert I\right\rangle ,$ and denoting $\left\vert \rho \right\rangle
=\rho \left\vert I\right\rangle ,$ and using (\ref{4}) we have the
time-evolution equation for $\left\vert \rho \left( t\right) \right\rangle ,$
\begin{equation}
\frac{d}{dt}\left\vert \rho \left( t\right) \right\rangle =\left[
\begin{array}{c}
g\left( 2a^{\dagger }\tilde{a}^{\dagger }-aa^{\dagger }-\tilde{a}\tilde{a}%
^{\dagger }\right) \\
+\kappa \left( 2a\tilde{a}-a^{\dagger }a-\tilde{a}^{\dagger }\tilde{a}\right)%
\end{array}%
\right] \left\vert \rho \left( t\right) \right\rangle .  \label{5}
\end{equation}%
where $\left\vert \rho _{0}\right\rangle \equiv \rho _{0}\left\vert
I\right\rangle ,$ $\rho _{0}$ is the initial density operator.

The formal solution of (\ref{5}) is
\begin{equation}
\left\vert \rho \left( t\right) \right\rangle =U\left( t\right) \left\vert
\rho _{0}\right\rangle ,  \label{6}
\end{equation}%
and%
\begin{equation}
U\left( t\right) =\exp \left[
\begin{array}{c}
gt\left( 2a^{\dagger }\widetilde{a}^{\dagger }-aa^{\dagger }-\widetilde{a}%
\widetilde{a}^{\dagger }\right) \\
+\kappa t\left( 2a\widetilde{a}-a^{\dagger }a-\widetilde{a}^{\dagger }%
\widetilde{a}\right)%
\end{array}%
\right] .  \label{7}
\end{equation}%
It challenges us how to disentangle the exponential operator $U\left(
t\right) .$ This reminds us of two theorems about the normally ordered
expansion of multimode bosonic exponential operators, which is helpful to
disentangle $U\left( t\right) $.

\section{Two Theorems}

In order to find the disentangled form of (\ref{7}) we employ two new
theorems about the normally ordered expansion of multimode bosonic
exponential operators$\ ^{\cite{R10,R11}}:$

\textbf{Theorem 1}: The multimode bosonic exponential operator $\exp
\mathcal{H},$ where $\mathcal{H=}\frac{1}{2}B\Gamma \widetilde{B},$ $B$\ is
defined by
\begin{eqnarray}
B &\equiv &\left( A^{\dagger }\text{ }A\right) \equiv \left( a_{1}^{\dagger }%
\text{ }a_{2}^{\dagger }\cdot \cdot \cdot \text{ }a_{n}^{\dagger }\text{ }%
a_{1}\text{\ }a_{2}\cdot \cdot \cdot \text{\ }a_{n}\right) ,  \label{8} \\
\text{ }\widetilde{B} &=&\binom{\widetilde{A}^{\dagger }}{\widetilde{A}},
\notag
\end{eqnarray}%
$\Gamma $ is a $2n\times 2n$ matrix, has its $n$-mode coherent state
representation:%
\begin{equation}
\begin{array}{c}
\exp \mathcal{H}=\sqrt{\det Q}\int \prod\limits_{i=1}^{n}\frac{d^{2}Z_{i}}{%
\pi }|\left(
\begin{array}{cc}
Q & -L \\
-N & P%
\end{array}%
\right) \binom{\widetilde{Z}}{\widetilde{Z}^{\ast }}\rangle \langle \binom{%
\widetilde{Z}}{\widetilde{Z}^{\ast }}|,%
\end{array}
\label{9}
\end{equation}%
where the $n$-mode coherent state is defined as%
\begin{eqnarray}
|\binom{\widetilde{Z}}{\widetilde{Z}^{\ast }}\rangle  &\equiv &\left\vert
Z\right\rangle =D\left( Z\right) \left\vert \vec{0}\right\rangle ,\text{\ \ }
\label{10} \\
\text{\ }D\left( Z\right)  &\equiv &\exp \{A^{\dagger }\widetilde{Z}-A%
\widetilde{Z}^{\ast }\},  \notag
\end{eqnarray}%
and%
\begin{equation}
\left(
\begin{array}{cc}
Q & L \\
N & P%
\end{array}%
\right) =\exp \{\Gamma \Pi \},\text{ }\Pi =\left(
\begin{array}{cc}
0 & -I_{n} \\
I_{n} & 0%
\end{array}%
\right) .  \label{11}
\end{equation}%
$I_{n}$ is the $n\times n$ unit matrix. $Q,L,N,P$ are $n\times n$ complex
matrices, $\left(
\begin{array}{cc}
Q & L \\
N & P%
\end{array}%
\right) \equiv M$ is a symplectic matrix, obeying
\begin{equation}
M\Pi \tilde{M}=\Pi ,\text{ }\Pi \tilde{M}\Pi =-M^{-1},  \label{12}
\end{equation}%
or
\begin{eqnarray}
Q\tilde{L} &=&L\tilde{Q},\text{ }Q\tilde{P}-L\tilde{N}=I,\text{ }  \label{13}
\\
\text{ }N\tilde{P} &=&P\tilde{N},\text{ }P\tilde{Q}-N\tilde{L}=I.  \notag
\end{eqnarray}

\textbf{Theorem 2}: By performing the integration in (\ref{9}) with the
technique of integration within an ordered product of operators $^{\cite%
{R11,R12}}$, we have%
\begin{equation}
\begin{array}{c}
\exp \mathcal{H}=\frac{1}{\sqrt{\det P}}\exp \{-\frac{1}{2}A^{\dagger
}(LP^{-1})\widetilde{A}^{\dagger }\} \\
\times \exp \{A^{\dagger }(\ln \widetilde{P}^{-1})\widetilde{A}\}\exp \{%
\frac{1}{2}A(P^{-1}N)\widetilde{A}\}.%
\end{array}
\label{14}
\end{equation}

\bigskip Now we first appeal to Theorem 1, so we should identify $U\left(
t\right) $ in (\ref{7}) as $\exp \mathcal{H}$, where $A=\left(
\begin{array}{cc}
\widetilde{a} & a%
\end{array}%
\right) $. After putting $U\left( t\right) $ into the following symmetrized
matrix form%
\begin{equation}
U\left( t\right) =e^{\left( \kappa -g\right) t}\exp \left[ \frac{1}{2}%
B\Gamma \widetilde{B}\right]  \label{15}
\end{equation}%
with $\Gamma $ being the symmetric matrix%
\begin{equation}
\Gamma =t\left(
\begin{array}{cc}
2gJ_{2} & -\left( g+\kappa \right) I_{2} \\
-\left( g+\kappa \right) I_{2} & 2\kappa J_{2}%
\end{array}%
\right)  \label{16}
\end{equation}%
here
\begin{equation}
I_{2}=\left(
\begin{array}{cc}
1 & 0 \\
0 & 1%
\end{array}%
\right) ,\text{ \ }J_{2}=\left(
\begin{array}{cc}
0 & 1 \\
1 & 0%
\end{array}%
\right) ,\text{ }J_{2}^{2}=I_{2},\text{ }  \label{17}
\end{equation}%
we then follow (\ref{11}) to calculate exp$\left( \Gamma \Pi \right) $ with%
\begin{equation}
\Gamma \Pi =t\left(
\begin{array}{cc}
-\left( g+\kappa \right) I_{2} & -2gJ_{2} \\
2\kappa J_{2} & \left( g+\kappa \right) I_{2}%
\end{array}%
\right)  \label{18}
\end{equation}%
therefore%
\begin{equation}
e^{\Gamma \Pi }\equiv \left(
\begin{array}{cc}
Q & L \\
N & P%
\end{array}%
\right)  \label{19}
\end{equation}%
with%
\begin{equation}
\begin{array}{c}
Q\equiv \frac{ge^{\left( \kappa -g\right) t}-\kappa e^{\left( g-\kappa
\right) t}}{g-\kappa }I_{2},\text{ }L\equiv \frac{g\left[ e^{\left( \kappa
-g\right) t}-e^{\left( g-\kappa \right) t}\right] }{g-\kappa }J_{2}, \\
N\equiv \frac{\kappa \left[ e^{\left( g-\kappa \right) t}-e^{\left( \kappa
-g\right) t}\right] }{g-\kappa }J_{2},\text{ }P\equiv \frac{ge^{\left(
g-\kappa \right) t}-\kappa e^{\left( \kappa -g\right) t}}{g-\kappa }I_{2}.%
\end{array}
\label{20}
\end{equation}%
Thus according to Theorems 1 and 2 we have%
\begin{equation}
\begin{array}{c}
U\left( t\right) =\frac{\kappa -g}{\kappa e^{-2\left( g-\kappa \right) t}-g}%
\exp \left[ \frac{g\left[ 1-e^{-2\left( \kappa -g\right) t}\right] }{\kappa
-ge^{-2\left( \kappa -g\right) t}}\widetilde{a}^{\dagger }a^{\dagger }\right]
\\
\times \exp \left[ \left( \widetilde{a}^{\dagger }\widetilde{a}+a^{\dagger
}a\right) \ln \frac{\left( \kappa -g\right) e^{-\left( \kappa -g\right) t}}{%
\kappa -ge^{-2\left( \kappa -g\right) t}}\right] \\
\times \exp \left[ \frac{\kappa \left[ 1-e^{-2\left( \kappa -g\right) t}%
\right] }{\kappa -ge^{-2\left( \kappa -g\right) t}}a\widetilde{a}\right] ,%
\end{array}
\label{21}
\end{equation}%
where we have used%
\begin{equation}
\begin{array}{c}
LP^{-1}=\frac{g\left[ 1-e^{-2\left( \kappa -g\right) t}\right] }{%
ge^{-2\left( \kappa -g\right) t}-\kappa }J_{2}, \\
P^{-1}N=\frac{\kappa \left[ e^{-2\left( \kappa -g\right) t}-1\right] }{%
ge^{-2\left( \kappa -g\right) t}-\kappa }J_{2}%
\end{array}
\label{22}
\end{equation}%
and%
\begin{equation}
\sqrt{\det P}\equiv \frac{ge^{\left( g-\kappa \right) t}-\kappa e^{\left(
\kappa -g\right) t}}{g-\kappa }.  \label{23}
\end{equation}%
writing%
\begin{eqnarray}
T_{1} &=&\frac{1-e^{-2\left( \kappa -g\right) t}}{\kappa -ge^{-2t\left(
\kappa -g\right) }},\text{ }T_{2}=\frac{\left( \kappa -g\right) e^{-\left(
\kappa -g\right) t}}{\kappa -ge^{-2t\left( \kappa -g\right) }},\text{ }
\label{24} \\
T_{3} &=&\frac{\kappa -g}{\kappa -ge^{-2t\left( \kappa -g\right) }}=1-gT_{1},
\notag
\end{eqnarray}%
and using (\ref{4}), (\ref{6}) becomes%
\begin{equation}
\begin{array}{c}
\left\vert \rho \left( t\right) \right\rangle =U\left( t\right) \left\vert
\rho _{0}\right\rangle \\
=T_{3}e^{gT_{1}a^{\dagger }\tilde{a}^{\dagger }}\colon e^{\left(
T_{2}-1\right) \left( \tilde{a}^{\dagger }\tilde{a}+a^{\dagger }a\right)
}\colon e^{\kappa T_{1}a\tilde{a}}\left\vert \rho _{0}\right\rangle \\
=\sum\limits_{i,j=0}^{\infty }T_{3}\frac{\kappa ^{i}g^{j}T_{1}^{i+j}}{%
i!j!T_{2}^{2j}}e^{a^{\dagger }a\ln T_{2}}a^{\dagger j}a^{i}\rho
_{0}a^{\dagger i}a^{j}e^{a^{\dagger }a\ln T_{2}}\left\vert \eta
=0\right\rangle ,%
\end{array}
\label{25}
\end{equation}%
or%
\begin{equation}
\rho \left( t\right) =\sum_{i,j=0}^{\infty }M_{ij}\rho _{0}M_{ij}^{\dagger },
\label{26}
\end{equation}%
where%
\begin{equation}
M_{ij}=\sqrt{\frac{\kappa ^{i}g^{j}T_{3}T_{1}^{i+j}}{i!j!T_{2}^{2j}}}%
e^{a^{\dagger }a\ln T_{2}}a^{\dagger j}a^{i}  \label{27}
\end{equation}%
is the Kraus operator, and one can check%
\begin{equation}
\sum_{i,j=0}^{\infty }M_{ij}^{\dagger }M_{ij}=1.  \label{28}
\end{equation}

\section{The Photon Number Evolution}

Now we have an explicit solution of the density matrix of a laser (\ref{26}%
), we first calculate the evolution of photon number in a laser process
initially in a pure coherent state, i.e., $\rho _{0}=\left\vert
z\right\rangle \left\langle z\right\vert $, $\ \left\vert z\right\rangle
=\exp [-|z|^{2}/2+za^{\dagger }]\left\vert 0\right\rangle ,$%
\begin{equation}
\begin{array}{c}
\left\langle n\right\rangle =Tr\left[ \rho \left( t\right) a^{\dagger }a%
\right] \\
=e^{\kappa T_{1}\left\vert z\right\vert ^{2}}Tr\left[ \sum\limits_{j=0}^{%
\infty }T_{3}\frac{g^{j}T_{1}^{j}}{j!T_{2}^{2j}}e^{a^{\dagger }a\ln
T_{2}}a^{\dagger j}\left\vert z\right\rangle \left\langle z\right\vert
a^{j}e^{a^{\dagger }a\ln T_{2}}a^{\dagger }a\right]%
\end{array}
\label{29}
\end{equation}%
Then using $\left\vert 0\right\rangle \left\langle 0\right\vert
=:e^{-a^{\dagger }a}:,$ the normal ordering form of the vacuum projector, as
well as $\int \frac{d^{2}z}{\pi }\left\vert z\right\rangle \left\langle
z\right\vert =1$ we have the expected photon number evolution formula
\begin{equation}
\begin{array}{c}
\left\langle n\right\rangle \\
=T_{3}e^{\left( \kappa T_{1}-1\right) \left\vert z\right\vert ^{2}}Tr\left[
\sum\limits_{j=0}^{\infty }\frac{g^{j}T_{1}^{j}}{j!}a^{\dagger
j}e^{zT_{2}a^{\dagger }}:e^{-a^{\dagger }a}:e^{z^{\ast
}T_{2}a}a^{j}a^{\dagger }a\right] \\
=T_{3}e^{\left( \kappa T_{1}-1\right) \left\vert z\right\vert ^{2}}Tr\left[
e^{zT_{2}a^{\dagger }}e^{a^{\dagger }a\ln \left( gT_{1}\right) }e^{z^{\ast
}T_{2}a}a^{\dagger }a\right] \\
=T_{3}e^{\left( \kappa T_{1}-1\right) \left\vert z\right\vert ^{2}}Tr\left[
e^{zT_{2}a^{\dagger }}\left( gT_{1}a^{\dagger }+z^{\ast }T_{2}\right)
e^{a^{\dagger }a\ln \left( gT_{1}\right) }e^{z^{\ast }T_{2}a}a\right] \\
=T_{3}e^{\left( \kappa T_{1}-1\right) \left\vert z\right\vert ^{2}}\int
\frac{d^{2}z^{\prime }}{\pi }\left\langle z^{\prime }\right\vert \\
\times :e^{zT_{2}a^{\dagger }+z^{\ast }T_{2}a+\left( gT_{1}-1\right)
a^{\dagger }a}\left( gT_{1}a^{\dagger }+z^{\ast }T_{2}\right) a:\left\vert
z^{\prime }\right\rangle \\
=g\frac{1-e^{-2\left( \kappa -g\right) t}}{\kappa -g}+\left\vert
z\right\vert ^{2}e^{-2\left( \kappa -g\right) t}.%
\end{array}
\label{30}
\end{equation}

We can easily write down the asymptotic behavior of $\left\langle
n\right\rangle $ when $t\rightarrow +\infty $ as the following:

If $\kappa =g$,$\ $then $\left\langle n\right\rangle =\left\vert
z\right\vert ^{2}+2gt$ , the photon number increases linearly with time.

\bigskip If $\kappa <g$, then $\left\langle n\right\rangle \sim \left( \frac{%
g}{g-\kappa }+\left\vert z\right\vert ^{2}\right) e^{2\left( g-\kappa
\right) t}$, the photon number increases exponentially when $t\rightarrow
+\infty .$

If $\kappa >g$, then $\left\langle n\right\rangle \sim \frac{g}{\kappa -g}$,
the expected photon number approaches a constant when $t\rightarrow +\infty
. $

\section{\protect\bigskip The Entropy Evolution in a Laser}

We now calculate how the entropy of a laser evolves with time. Using (\ref%
{26}), the density matrix $\rho \left( t\right) $ of a laser initially in a
pure coherent state $\left\vert z\right\rangle \left\langle z\right\vert $ is%
\begin{equation}
\begin{array}{c}
\rho \left( t\right) =T_{3}\exp \left[ \left\vert z\right\vert
^{2}e^{2\left( g-\kappa \right) t}\ln gT_{1}\right] \\
\times \sum\limits_{j=0}^{\infty }\frac{g^{j}T_{1}^{j}}{j!T_{2}^{2j}}%
:a^{\dagger j}a^{j}e^{za^{\dagger }+z^{\ast }a-a^{\dagger }a}:e^{a^{\dagger
}a\ln T_{2}} \\
=T_{3}e^{\kappa T_{1}\left\vert z\right\vert ^{2}-\left\vert z\right\vert
^{2}}e^{zT_{2}a^{\dagger }}e^{a^{\dagger }a\ln \left( gT_{1}\right)
}e^{z^{\ast }T_{2}a}.%
\end{array}
\label{31}
\end{equation}%
By the Baker--Campbell--Hausdorff formula, if%
\begin{equation}
\left[ X,Y\right] =\lambda Y+\mu ,  \label{32}
\end{equation}%
then
\begin{equation}
\exp X\exp Y=\exp \left( X+\frac{\lambda Y+\mu }{1-e^{-\lambda }}-\frac{\mu
}{\lambda }\right) .  \label{33}
\end{equation}%
we can compact the three exponentials in (\ref{31}) into a single
exponential
\begin{equation}
\begin{array}{c}
\rho \left( t\right) =T_{3}\exp \left[ \left\vert z\right\vert
^{2}e^{2\left( g-\kappa \right) t}\ln gT_{1}\right] \\
\times \exp \left\{ \left[ a^{\dagger }a-e^{\left( g-\kappa \right) t}\left(
za^{\dagger }+z^{\ast }a\right) \right] \ln gT_{1}\right\} .%
\end{array}
\label{34}
\end{equation}%
with direct calculations. Thus we see how a pure state $\left\vert
z\right\rangle \left\langle z\right\vert $ evolves into a mixed state, so
the entangled state representation in (\ref{3}) can well expose the
entanglement between the system and its environment. Then the logarithm of $%
\rho \left( t\right) $ can be evaluated as
\begin{equation}
\begin{array}{c}
\ln \rho \left( t\right) =\ln T_{3}+\left\vert z\right\vert ^{2}e^{2\left(
g-\kappa \right) t}\ln gT_{1} \\
+\left[ a^{\dagger }a-e^{\left( g-\kappa \right) t}\left( za^{\dagger
}+z^{\ast }a\right) \right] \ln gT_{1}.%
\end{array}
\label{35}
\end{equation}%
Therefore the von-Neumann entropy of $\rho \left( t\right) $ is%
\begin{equation}
\begin{array}{c}
S\left( \rho \left( t\right) \right) /k_{B}=-Tr\left[ \rho \ln \rho \right]
\\
=-Tr\left[ \rho \left( \ln T_{3}+\left\vert z\right\vert ^{2}e^{2\left(
g-\kappa \right) t}\ln gT_{1}\right) \right] -T_{3}e^{\left( \kappa
T_{1}-1\right) \left\vert z\right\vert ^{2}}\ln gT_{1} \\
\times Tr\left[ e^{zT_{2}a^{\dagger }}e^{a^{\dagger }a\ln gT_{1}}e^{z^{\ast
}T_{2}a}\left( a^{\dagger }a-e^{\left( g-\kappa \right) t}\left( za^{\dagger
}+z^{\ast }a\right) \right) \right] \\
=-\ln T_{3}-\left\vert z\right\vert ^{2}e^{2\left( g-\kappa \right) t}\ln
gT_{1}-T_{3}e^{\left( \kappa T_{1}-1\right) \left\vert z\right\vert ^{2}}\ln
gT_{1} \\
\times Tr\left[ e^{zT_{2}a^{\dagger }}e^{a^{\dagger }a\ln gT_{1}}e^{z^{\ast
}T_{2}a}\left( a^{\dagger }a-e^{\left( g-\kappa \right) t}\left( za^{\dagger
}+z^{\ast }a\right) \right) \right] ,%
\end{array}
\label{36}
\end{equation}%
where $k_{B}$ is the Boltzmann constant. Since%
\begin{equation}
\begin{array}{c}
e^{zT_{2}a^{\dagger }}e^{a^{\dagger }a\ln gT_{1}}e^{z^{\ast }T_{2}a}\left[
a^{\dagger }a-e^{\left( g-\kappa \right) t}za^{\dagger }\right] \\
=e^{zT_{2}a^{\dagger }}e^{a^{\dagger }a\ln gT_{1}}\left( a^{\dagger
}+z^{\ast }T_{2}\right) e^{z^{\ast }T_{2}a}a \\
-e^{\left( g-\kappa \right) t}ze^{zT_{2}a^{\dagger }}e^{a^{\dagger }a\ln
gT_{1}}\left( a^{\dagger }+z^{\ast }T_{2}\right) e^{z^{\ast }T_{2}a} \\
=e^{zT_{2}a^{\dagger }}\left( gT_{1}a^{\dagger }+z^{\ast }T_{2}\right)
e^{a^{\dagger }a\ln gT_{1}}e^{z^{\ast }T_{2}a}a \\
-e^{\left( g-\kappa \right) t}ze^{zT_{2}a^{\dagger }}\left( gT_{1}a^{\dagger
}+z^{\ast }T_{2}\right) e^{a^{\dagger }a\ln gT_{1}}e^{z^{\ast }T_{2}a},%
\end{array}
\label{37}
\end{equation}%
therefore%
\begin{equation}
\begin{array}{c}
Tr\left[ e^{zT_{2}a^{\dagger }}e^{a^{\dagger }a\ln gT_{1}}e^{z^{\ast
}T_{2}az^{\ast }T_{2}a}\left( a^{\dagger }a-e^{\left( g-\kappa \right)
t}\left( za^{\dagger }+z^{\ast }a\right) \right) \right] \\
=\int \frac{d^{2}z^{\prime }}{\pi }\left\langle z^{\prime }\right\vert
:e^{zT_{2}a^{\dagger }+z^{\ast }T_{2}a+\left( gT_{1}-1\right) a^{\dagger }a}
\\
\times \left[ \left( gT_{1}a^{\dagger }+z^{\ast }T_{2}\right) \left(
a-e^{\left( g-\kappa \right) t}z\right) -e^{\left( g-\kappa \right)
t}z^{\ast }a\right] :\left\vert z^{\prime }\right\rangle \\
=\int \frac{d^{2}z^{\prime }}{\pi }e^{zT_{2}z^{\prime \ast }+z^{\ast
}T_{2}z^{\prime }+\left( gT_{1}-1\right) \left\vert z^{\prime }\right\vert
^{2}} \\
\times \left[ \left( gT_{1}z^{\prime \ast }+z^{\ast }T_{2}\right) \left(
z^{\prime }-e^{\left( g-\kappa \right) t}z\right) -e^{\left( g-\kappa
\right) t}z^{\ast }z^{\prime }\right] .%
\end{array}
\label{38}
\end{equation}%
Finally the entropy variation law is%
\begin{equation}
S\left( \rho \left( t\right) \right) =-k_{B}\left( \ln T_{3}+\frac{gT_{1}\ln
gT_{1}}{1-gT_{1}}\right)  \label{39}
\end{equation}

\begin{figure}[b]
\vspace{0.9in} \includegraphics{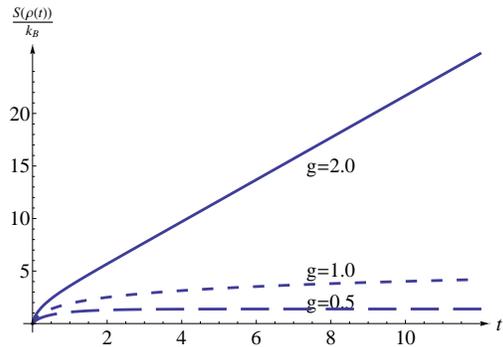}
\caption{{\protect\small $S\left( \protect\rho \left( t\right) \right)
/k_{B} $ for $z=4,\protect\kappa =1$ and $g=2,1,0.5$ respectively}}
\label{f1}
\end{figure}
\newpage

\bigskip We also write down the asymptotic behavior of $S\left( \rho \left(
t\right) \right) $ when $t\rightarrow +\infty $ as the following:

If $\kappa =g$, then $S\left( \rho \left( t\right) \right) /k_{B}\sim $ $%
1+\ln \left( 2gt\right) $ as $t\rightarrow +\infty $, the entropy increases
logarithmically.

If $\kappa <g$ then $S\left( \rho \left( t\right) \right) /k_{B}\sim 1+\ln
\frac{g}{g-\kappa }+2\left( g-\kappa \right) t$ as $t\rightarrow +\infty $,
the entropy increases linearly.

If $\kappa >g$ then $S\left( \rho \left( t\right) \right) /k_{B}\sim \ln
\frac{\kappa }{\kappa -g}+\frac{g}{\kappa -g}\ln \frac{\kappa }{g}$ as $%
t\rightarrow +\infty $, the entropy approaches a constant.

The results of expected\ photon number and entropy of the laser do not\
depend on the phase of parameter $z$, as one should expect, since the
absolute phase of $z$ in a coherent state is non-physical. It is remarkable
that the entropy is completely independent of $z$.

Plots of $S\left( \rho \left( t\right) \right) $ and the "specific entropy" $%
\frac{S\left( \rho \left( t\right) \right) }{\left\langle n\right\rangle }$
in unit of $k_{B}$ for $z=4$ (16 photons in average)$,$ $\kappa =1$ and $%
g=2,1,0.5$ respectively are shown in Figure 1 and 2. Besides the We can see
clearly from the two figures that when the pumping rate $g$ is less than the
loss rate $\kappa $, the photon number and entropy will approach to
constants, the photons are in fact in sort of thermo-equilibrium with an
equivalent temperature $T=\frac{\hbar \omega }{k_{B}}\ln \frac{\kappa }{g}$.
When $g$ is larger than $\kappa $, while the entropy increases linearly with
time, the expected number of photons increases much more fast, therefore the
specific entropy will goes to zero exponentially. The photons in the laser
are highly coherent in this case. The above results indeicate that a laser
can generate laser beam only if it works with sufficiently high pumping rate
$g$.

\begin{figure}[t!]
\vspace{0.9in} \includegraphics{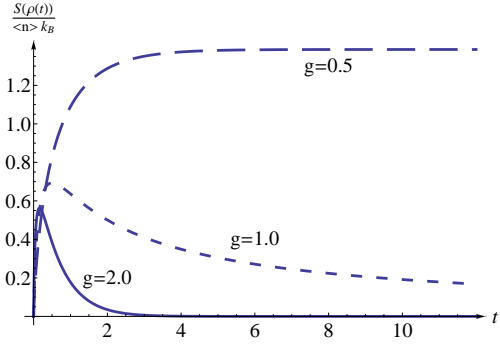}
\caption{{\protect\small $S\left( \protect\rho \left( t\right) \right) /{%
(\langle n\rangle k_{B})}$ for $z=4,\protect\kappa =1$ and $g=2,1,0.5$
respectively}}
\label{f2}
\end{figure}

\end{document}